\documentclass[aps,prd,twocolumn,superscriptaddress,showpacs,nofootinbib]{revtex4-1}

\usepackage{amsmath,amssymb,graphicx}
\usepackage{color}
\usepackage{pslatex}
\usepackage{graphicx}
\usepackage{psfrag}

\usepackage{graphicx}
\usepackage{hyperref}

\usepackage[normalem]{ulem}
\usepackage{microtype}

\begin{document}

\title{Connecting the new H.E.S.S. diffuse emission at the Galactic center with the \textit{Fermi} GeV excess: a combination of millisecond pulsars and heavy dark matter?}

\author{Thomas Lacroix}
\affiliation{Institut d'Astrophysique de Paris, UMR 7095, CNRS, UPMC Universit\'{e} Paris 6, Sorbonne Universit\'{e}s, 98 bis boulevard Arago, 75014 Paris, France}
\affiliation{Laboratoire Univers \& Particules de Montpellier (LUPM), CNRS \& Universit\'{e} de Montpellier (UMR-5299), Place Eug\`{e}ne Bataillon, F-34095 Montpellier Cedex 05, France}
\author{Joseph Silk}
\affiliation{Institut d'Astrophysique de Paris, UMR 7095, CNRS, UPMC Universit\'{e} Paris 6, Sorbonne Universit\'{e}s, 98 bis boulevard Arago, 75014 Paris, France}
\affiliation{AIM-Paris-Saclay, CEA/DRF/Irfu, CNRS, Univ. Paris VII, F-91191 Gif-sur-Yvette, France}
\affiliation{The Johns Hopkins University, Department of Physics and Astronomy,
3400 N. Charles Street, Baltimore, Maryland 21218, USA}
\affiliation{Beecroft Institute of Particle Astrophysics and Cosmology, Department of Physics,
University of Oxford, Denys Wilkinson Building, 1 Keble Road, Oxford OX1 3RH, United Kingdom}
\author{Emmanuel Moulin}
\affiliation{DRF/Irfu, Service de Physique des Particules, CEA Saclay, F-91191 Gif-Sur-Yvette Cedex, France}
\author{C\'{e}line B\oe hm}
\affiliation{Institute for Particle Physics Phenomenology, Durham University, Durham, DH1 3LE, United Kingdom}
\affiliation{LAPTH, Universit\'{e} de Savoie, CNRS, BP 110, 74941 Annecy-Le-Vieux, France}

\date{\today}

\begin{abstract}
The H.E.S.S. collaboration has reported a high-energy spherically symmetric diffuse $\gamma$-ray emission in the inner 50 pc of the Milky Way, up to $\sim 50\ \mathrm{TeV}$. Here we propose a leptonic model which provides an alternative to the hadronic scenario presented by the H.E.S.S. collaboration, and connects the newly reported TeV emission to the \textit{Fermi}-LAT Galactic center GeV excess. Our model relies on a combination of inverse Compton emission from a population of millisecond pulsars---which can account for the GeV excess---and a supermassive black hole-induced spike of heavy ($\sim 60\ \mathrm{TeV}$) dark matter particles annihilating into electrons with a sub-thermal cross-section. With an up-to-date interstellar radiation field, as well as a standard magnetic field and diffusion set-up, our model accounts for the spectral morphology of the detected emission. Moreover, we show that the dark matter induced emission reproduces the spatial morphology of the H.E.S.S. signal above $\sim 10\ \mathrm{TeV}$, while we obtain a slightly more extended component from pulsars at lower energies, which could be used as a prediction for future H.E.S.S. observations. 
\end{abstract}


\maketitle

\section{Introduction}
\label{Introduction}

The H.E.S.S. collaboration has released the most detailed high-energy $\gamma$-ray view to date of the inner 300 pc of the Galactic center (GC) region, thanks to improved statistics accumulated from 10 years of observation of the GC. In addition to the previously observed Galactic ridge emission~\cite{Aharonian2006a}, a spherically symmetric diffuse emission has been detected between $\sim 200$ GeV and 50 TeV in the inner 50 pc \cite{HESSextended2016}. Specifically, this emission has been extracted in an open ring centered on the GC, with azimuthal size 294 deg, and inner and outer radii 0.15 and 0.45 deg, respectively, as shown in Fig.~1 of Ref.~\cite{HESSextended2016}. This corresponds to a solid angle of $\Delta\Omega =1.4 \times 10^{-4}\ \rm sr$. Here we focus on this new feature, hereafter referred to as the H.E.S.S. diffuse emission, which is distinct, both in spatial and spectral morphologies, from the central source HESS J1745-290, which has an angular size of 0.1 deg. 

In the hadronic scenario described in Ref.~\cite{HESSextended2016}, TeV $\gamma$-rays originate from the decay of neutral pions produced by collisions of protons accelerated by the central black hole (BH) Sgr A* with ambient gas. In this work, we explore an alternative leptonic interpretation which relates the H.E.S.S. diffuse emission detected in the inner 50 pc to the excess of GeV $\gamma$-rays at the GC. The latter was reported in the data of the \textit{Fermi} Large Area Telescope (LAT) by several independent groups \cite{Goodenough2009,Vitale2009,Hooper2011,Hooper2011a,Abazajian2012,Abazajian2013,Gordon2013,Abazajian2014,Daylan2014,Calore2015a,Ajello2015}, and may originate from millisecond pulsars (MSPs), as pointed out in Refs.~\cite{Abazajian2012,Gordon2013,Macias2014,Abazajian2011,Wharton2012,Mirabal2013,Yuan2014,Petrovic2015}, and reinforced by several more recent papers \cite{Bartels2015,Brandt2015,Lee2015}. Here we show that such a population of MSPs may also contribute to the observed diffuse emission. More specifically, electrons in the pulsar winds can be accelerated to energies of a few tens of TeV and thus significantly contribute to the H.E.S.S. diffuse emission via inverse Compton scattering off ambient radiation fields. 

Our leptonic model of the diffuse emission evades the constraints discussed in Ref.~\cite{HESSextended2016}. In particular, the model must account for the hardness of the observed spectrum, and the propagation set-up must allow electrons to diffuse to sufficiently large distances. The MSP component turns out to be insufficient to account for the whole emission, and an additional harder component is needed. This motivates us to consider a multi-TeV DM candidate, which would actually produce $\gamma$-rays in the energy range of interest. As discussed in the following, the DM density profile must be strongly contracted in the very inner region in order for the associated $\gamma$-ray flux from DM annihilations to contribute significantly to the H.E.S.S. emission. A supermassive BH-induced density spike \cite{spikeGS} would provide the required enhancement of the DM annihilation signal in $\gamma$-rays. 

Therefore, in this paper, we show that a combination of MSPs that account for the GeV excess at the GC, and a SMBH-induced spike of heavy DM can explain the H.E.S.S. diffuse emission in the inner 50 pc.\footnote{In the region of the H.E.S.S. diffuse signal lies the massive molecular cloud Sgr C \cite{HESSextended2016}. A $\gamma$-ray contamination to the diffuse signal, from a source located in this cloud, cannot be excluded. Our model of the overall diffuse signal in terms of a combination of MSPs and heavy DM should therefore be interpreted as an upper limit.} Section~\ref{Theory} provides a short description of the TeV $\gamma$-ray emission expected for pulsars and dark matter annihilations. In section~\ref{Results}, we show our results on the modeling of the H.E.S.S. diffuse emission. We conclude in section~\ref{Conclusion}.

\section{Models of the TeV gamma-ray emission}
\label{Theory}
\subsection{Inverse Compton emission from MSPs}

The rotation energy of MSPs has been shown to power a high-energy $e^{\pm}$ wind \cite{Rees1974}. The interaction of this pulsar wind with the interstellar medium may create a shock which can accelerate $e^{\pm}$ to very high energies \cite{Bednarek:2007nn,Bednarek2013,Yuan2015}. Their maximum energy $E_{\mathrm{max}}$ is limited by their ability to escape the shock region, and by their synchrotron losses. This energy can be as high as a few tens of TeV, potentially up to 100 TeV \cite{Bednarek2013}. The resulting $e^{\pm}$ injection spectrum follows a power law, with the maximum energy accounted for by an exponential cut-off \cite{Bednarek2013}, and is expressed as $\left. \mathrm{d}N_{\mathrm{e}}/\mathrm{d}E_{\mathrm{inj}} \right | _{\mathrm{MSP}} \propto E_{\mathrm{inj}}^{-2} \exp (-E_{\mathrm{inj}}/E_{\mathrm{max}})$. 

These $e^{\pm}$ emit high-energy $\gamma$-rays by upscattering photons of the interstellar radiation field (ISRF) via the inverse Compton (IC) process. The resulting $\gamma$-ray spectrum can extend to very high energies, up to the range of interest for H.E.S.S. observations. This led the authors of Refs.~\cite{Bednarek2013,Yuan2015} to claim that IC emission from MSPs could be responsible for the H.E.S.S. central source data, based on a spectral analysis. In principle, $e^{\pm}$ also emit bremsstrahlung by interacting with nuclei of the ambient gas, but this component is only relevant in the GeV range and is negligible with respect to IC over the energy range of interest here.

In our model, the spatial distribution of MSPs is fixed by the \textit{Fermi}-LAT GeV excess data. There is however uncertainty on the fraction $f_{e^{\pm}}$ of spin-down power released by pulsars in the electron wind (which impacts the normalization of the subsequent $\gamma$-ray flux) and the maximum electron energy. The normalization of the $e^{\pm}$  injection spectrum (which can be absorbed in $f_{e^\pm}$) is worked out by fitting the spectrum of the H.E.S.S. diffuse emission, and we check the consistency of the result as discussed in the following.

The MSP-induced IC flux is obtained by integrating the emissivity $j_{\mathrm{MSP}}$ over the line of sight (l.o.s.) coordinate $s$ and the field of view (fov) $\Delta\Omega$ \citep{GeV_excess_my_paper}:
\begin{align}
\label{secondary_flux_MSP}
\left. E_{\gamma}^{2}\dfrac{\mathrm{d}n}{\mathrm{d}E_{\gamma}} \right | _{\mathrm{MSP}}^{\mathrm{IC}} & = \int_{\Delta\Omega} \left. E_{\gamma}^{2}\dfrac{\mathrm{d}n}{\mathrm{d}E_{\gamma}\mathrm{d}\Omega} \right | _{\mathrm{MSP}}^{\mathrm{IC}} \, \mathrm{d}\Omega \nonumber \\
& = \dfrac{E_{\gamma}}{4\pi} \int_{\Delta\Omega} \! \int_{\mathrm{l.o.s.}} \! j_{\mathrm{MSP}}(E_{\gamma},\vec{x}) \, \mathrm{d}s \, \mathrm{d}\Omega,
\end{align}
with the emissivity given by the convolution of the MSP $e^{\pm}$ spectrum after propagation $\psi_{\mathrm{MSP}}$ and the IC emission spectrum $P_{\mathrm{IC,G}}$ (see e.g.~Ref.~\cite{Cirelli_cookbook}; the G subscript refers to the standard ISRF implemented in the GALPROP code \footnote{\url{http://galprop.stanford.edu/}}):
\begin{equation}
j_{\mathrm{MSP}}(E_{\gamma},\vec{x}) = \int_{E_{\gamma}}^{E_{\mathrm{max}}} \! P_{\mathrm{IC,G}} (E_{\gamma},E_{\mathrm{e}},\vec{x}) \psi_{\mathrm{MSP}}(E_{\mathrm{e}},\vec{x}) \, \mathrm{d}E_{\mathrm{e}}.
\end{equation}
The integral over solid angle in Eq.~\eqref{secondary_flux_MSP} is performed over the fov of the H.E.S.S. diffuse emission, $\Delta\Omega = 1.4 \times 10^{-4}\ \rm sr$. 

The $e^{\pm}$ spectrum from MSPs after propagation $\psi_{\mathrm{MSP}}$, in a steady state and accounting for energy losses and spatial diffusion, reads (see e.g.~Refs.~\cite{3D_paper,Delahaye2010})
\begin{equation}
\label{electron_spectrum_MSP}
\psi_{\mathrm{MSP}} (E,\vec{x}) \propto \dfrac{1}{b_{\mathrm{G}}(E,\vec{x})} \int_{E}^{E_{\mathrm{max}}} \! \tilde{I}_{\vec{x},\mathrm{MSP}}(E,E_{\mathrm{inj}}) \left. \dfrac{\mathrm{d}N_{\mathrm{e}}}{\mathrm{d}E_{\mathrm{inj}}} \right | _{\mathrm{MSP}} \, \mathrm{d}E_{\mathrm{inj}},
\end{equation}
where $b_{\mathrm{G}}(\vec{x},E)$ is the sum of the synchrotron and IC loss rates, corresponding to the GALPROP losses tabulated in Ref.~\cite{Cirelli_cookbook_secondaries}. Assuming a 10 $\mu$G magnetic field in the Galactic center region, the synchrotron losses are comparable to IC losses for TeV electrons.\footnote{Our value of the magnetic field is compatible with the values used in the GALPROP code in the inner 50 pc region.} 

The \textit{halo function} $\tilde{I}_{\vec{x},\mathrm{MSP}}$ accounts for spatial diffusion of electrons injected according to the MSP profile, and is computed exactly as in Refs.~\cite{Spike_GC_my_paper,GeV_excess_my_paper}. The diffusion coefficient is parametrized as $K(E) = K_{0} \left( E/E_{0} \right) ^{\delta}$, with $K_{0} = 6.67 \times 10^{26} \ \rm cm^2\ s^{-1}$, $E_{0} = 1\ \rm GeV$, $\delta = 0.7$, and a half height of $L = 4\ \mathrm{kpc}$ for the diffusion zone. The normalization $K_{0}$ is determined so that electrons of a few tens of TeV can diffuse on a scale of order a few 10 pc with the prescription $\lambda_{\rm D}/t_{\rm loss} \le c$---where the characteristic loss time $t_{\rm loss} = E/b(E)$ is computed for consistency for the enhanced IC losses relevant for electrons injected in a DM spike as discussed in the next section---, in agreement with the requirement $R^2/6K \ge R/c$, with $R$ the size of the region ($\sim$ 50 pc). Our diffusion parameters are consistent with other parametrizations used in the literature (see for instance Ref.~\cite{Lavalle:2014kca}). Although the recent AMS02 data on $\bar{p}/p$~\cite{Giesen:2015ufa} and B/C \cite{Genolini:2015cta} ratios tend to favor a relatively milder energy dependence of the diffusion coefficient in the Galactic disk, it is premature to draw any firm conclusions. On scales below 100 pc towards the central region of the Galaxy, the diffusion coefficient is even more uncertain (see for instance Ref.~\cite{Regis:2008ij}). For the density that enters into the calculation of $\tilde{I}_{\vec{x},\mathrm{MSP}}$, we use a squared generalized Navarro-Frenk-White (NFW) profile with slope $\sim 1.2$, consistent with the spatial morphology of the GeV excess, as found e.g.~in Ref.~\cite{Gordon2013}.

\begin{figure}[th!]
\centering
\includegraphics[width=\linewidth]{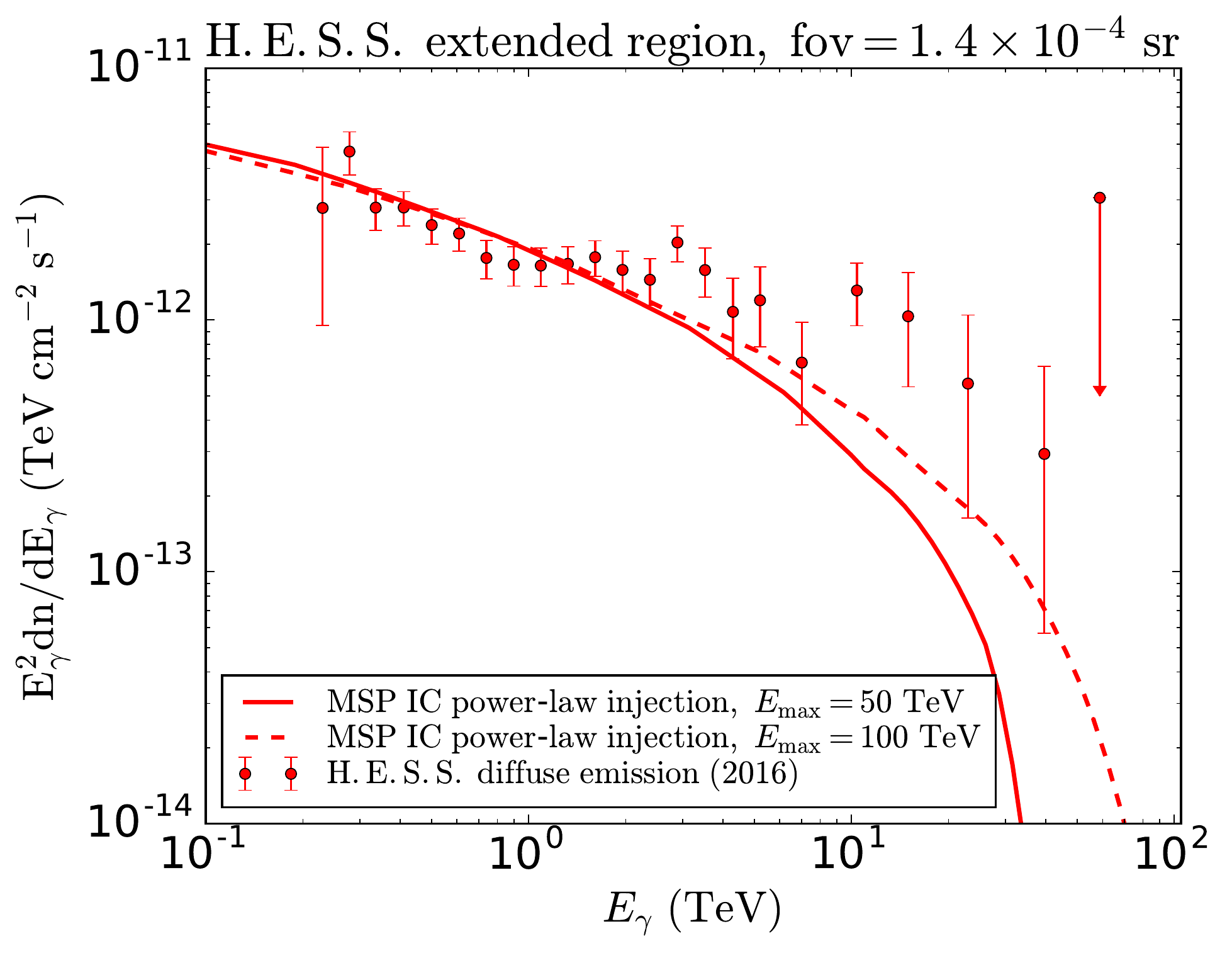}
\caption{$\gamma$-ray spectrum of the H.E.S.S. diffuse emission in the inner 50 pc of the GC for a fov of 
$\Delta\Omega_{\mathrm{d}}$ =1.4$\times$10$^{-4}$ sr. The IC $\gamma$-ray emission from MSPs is depicted for an energy cut-off  of 50 TeV (red solid line) and 100 TeV (red dashed line), and corresponds to $f_{e^{\pm}} \approx 0.1$.\label{extended_MSPs}}
\end{figure}

Figure~\ref{extended_MSPs} shows the predicted  $\gamma$-ray fluxes from the MSP model for maximum energies $E_{\mathrm{max}}$ of 50 TeV (solid) and 100 TeV (dashed). The best-fit model corresponds to a fraction $f_{e^{\pm}} \approx 0.1$.\footnote{A value of 0.1 for $f_{e^{\pm}}$ is actually well motivated since it actually corresponds to an electron wind power equal to the luminosity of the direct pulsar $\gamma$-ray emission that can account for the GeV excess \cite{Yuan2015}. We also note that a higher value of $f_{e^{\pm}}$ would overshoot the low energy part of the H.E.S.S. diffuse emission.} Figure~\ref{extended_MSPs} shows that the MSP emission can account for the H.E.S.S. diffuse emission up $\sim 10\ \mathrm{TeV}$ via IC scattering of electrons off the ISRF. However, even for an energy cut-off at 100 TeV, the MSP-induced IC emission fails to reproduce the high energy part of the spectrum beyond $\sim 10$ TeV. This is due to synchroton emission taking over IC emission above $\sim 10\ \mathrm{TeV}$, inducing a softening in the $\gamma$-ray spectrum. Therefore, an additional hard component is needed, and this provides the motivation for considering a contribution from annihilating DM.

\subsection{Annihilation signal from DM}
\label{DM_annihilation}

The DM component must satisfy several constraints. First, the DM candidate must feature an annihilation cross section smaller than $\sim 10^{-25}\ \mathrm{cm^{3}\ s^{-1}}$ at TeV masses to avoid tensions with recent observations \cite{Abramowski2011a}. Moreover, the DM density must be high enough in the GC region to produce a sufficiently high $\gamma$-ray flux. For these reasons, a regular NFW profile cannot account for the observed emission.

Therefore, we need to assume that a supermassive BH-induced spike---i.e.~a strong enhancement of the density---is present in the inner part of the DM density profile, following the prescription of Ref.~\cite{spikeGS}. More specifically, a spike is predicted to arise around a supermassive black hole (SMBH) growing adiabatically at the center of a DM halo, and typically corresponds to a density going as $r^{-\gamma_{\mathrm{sp}}}$, with $\gamma_{\mathrm{sp}} = 7/3$, below a parsec-scale radius $R_{\mathrm{sp}}$. We normalize the DM profile following Ref.~\cite{Spike_GC_my_paper}. The existence of a spike is actually debated, as discussed e.g.~in Refs.~\cite{Ullio2001,Gnedin2004}. In particular, stellar heating of DM particles may lead to a smoother profile instead---possibly down to a 1.5 slope---, and a spike may be destroyed by BH mergers, although current simulations used to model mergers do not have enough resolution to account for sub-parsec processes which are of crucial importance for the formation of spikes. Moreover, there is compelling evidence for a unique major merger involving the Milky Way about 12 billion years ago that led to the formation of the bulge \cite{Gilmore2001}, which would not have affected the survival of a spike. Ultimately, dedicated numerical simulations and observations are needed to settle the question.

\subsubsection{Prompt emission from DM annihilation}
The prompt $\gamma$-ray flux for an annihilation channel $f$ reads as 
\begin{align}
\label{prompt}
\left. E_{\gamma}^{2}\dfrac{\mathrm{d}n_{f}}{\mathrm{d}E_{\gamma}} \right | _{\mathrm{DM}} ^{\mathrm{prompt}} = \dfrac{E_{\gamma}^{2}}{4\pi} \left( \dfrac{\rho_{\odot}}{m_{\mathrm{DM}}}\right) ^{2} \dfrac{\left\langle \sigma v \right\rangle_{f}}{2} \dfrac{\mathrm{d}N_{\gamma,f}}{\mathrm{d}E_{\gamma}} \nonumber \\ \times \int_{\Delta\Omega_{\mathrm{c}}} \! \int_{\mathrm{l.o.s.}} \! \left( \dfrac{\rho(\vec{x})}{\rho_{\odot}}\right) ^{2} \, \mathrm{d}s \, \mathrm{d}\Omega,
\end{align}
with $\rho(\vec{x})$ the DM density at position $\vec{x}$, and $\rho_{\odot}$ the DM density in the solar neighborhood, which we take equal to $0.3\ \mathrm{GeV\ cm^{-3}}$ (see e.g.~\cite{Bovy2012}). $m_{\mathrm{DM}}$ is the DM mass, $\left\langle \sigma v \right\rangle _{f}$ the annihilation cross-section into the channel $f$ and $\mathrm{d}N_{\gamma,f}/\mathrm{d}E_{\gamma}$ the $\gamma$-ray spectrum from this final state, extracted from Ref.~\cite{Cirelli_cookbook}. In practice, the DM profile is so steep that the integral depends very weakly on the precise value of the fov $\Delta\Omega_{\mathrm{c}}$ as pointed out in Ref.~\cite{M87_limits_my_paper}, which we take equal to $10^{-5}\ \mathrm{sr}$, the size of the central source HESS J1745-290.

\begin{figure*}[th!]
\centering
\includegraphics[width=0.49\linewidth]{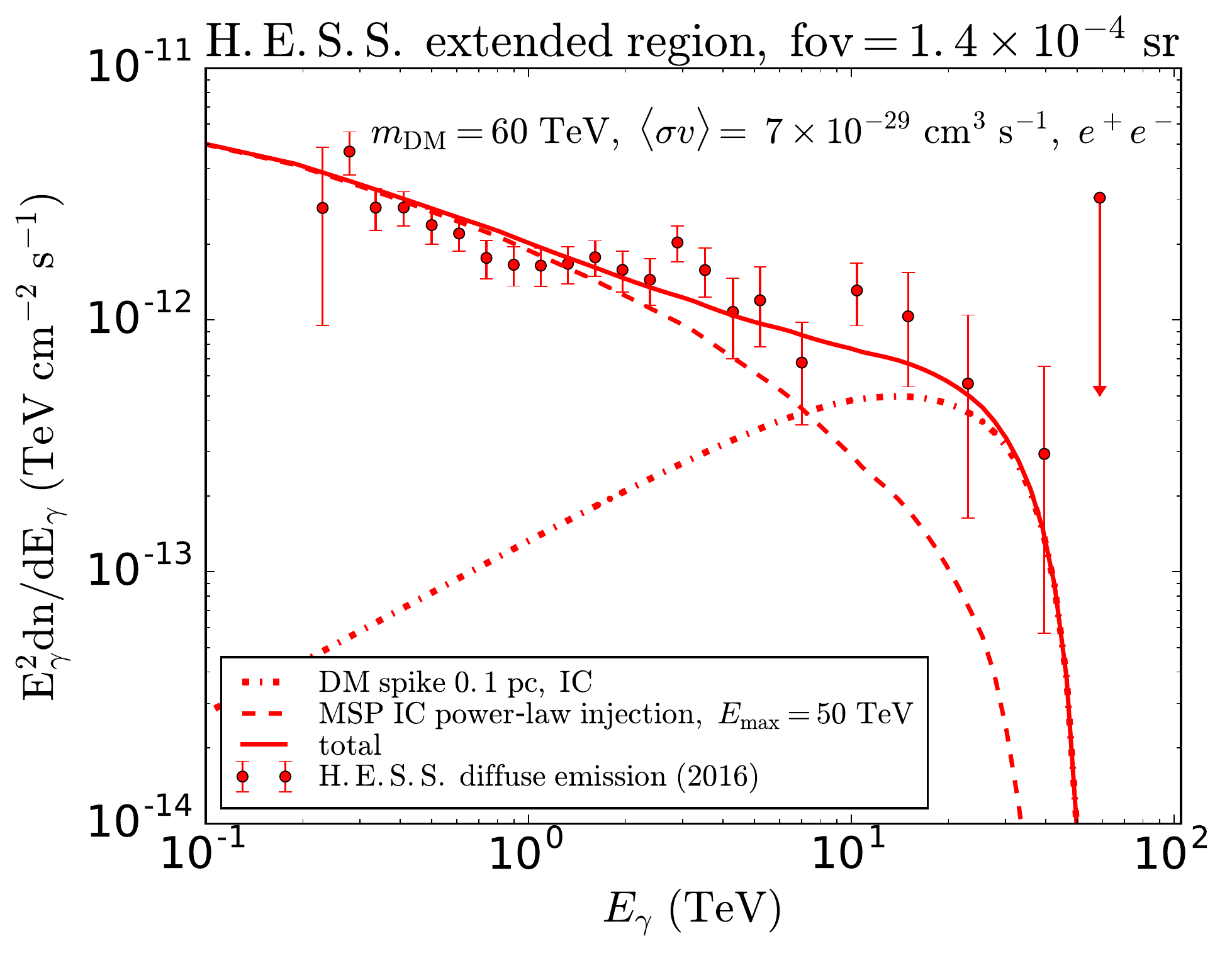}\includegraphics[width=0.49\linewidth]{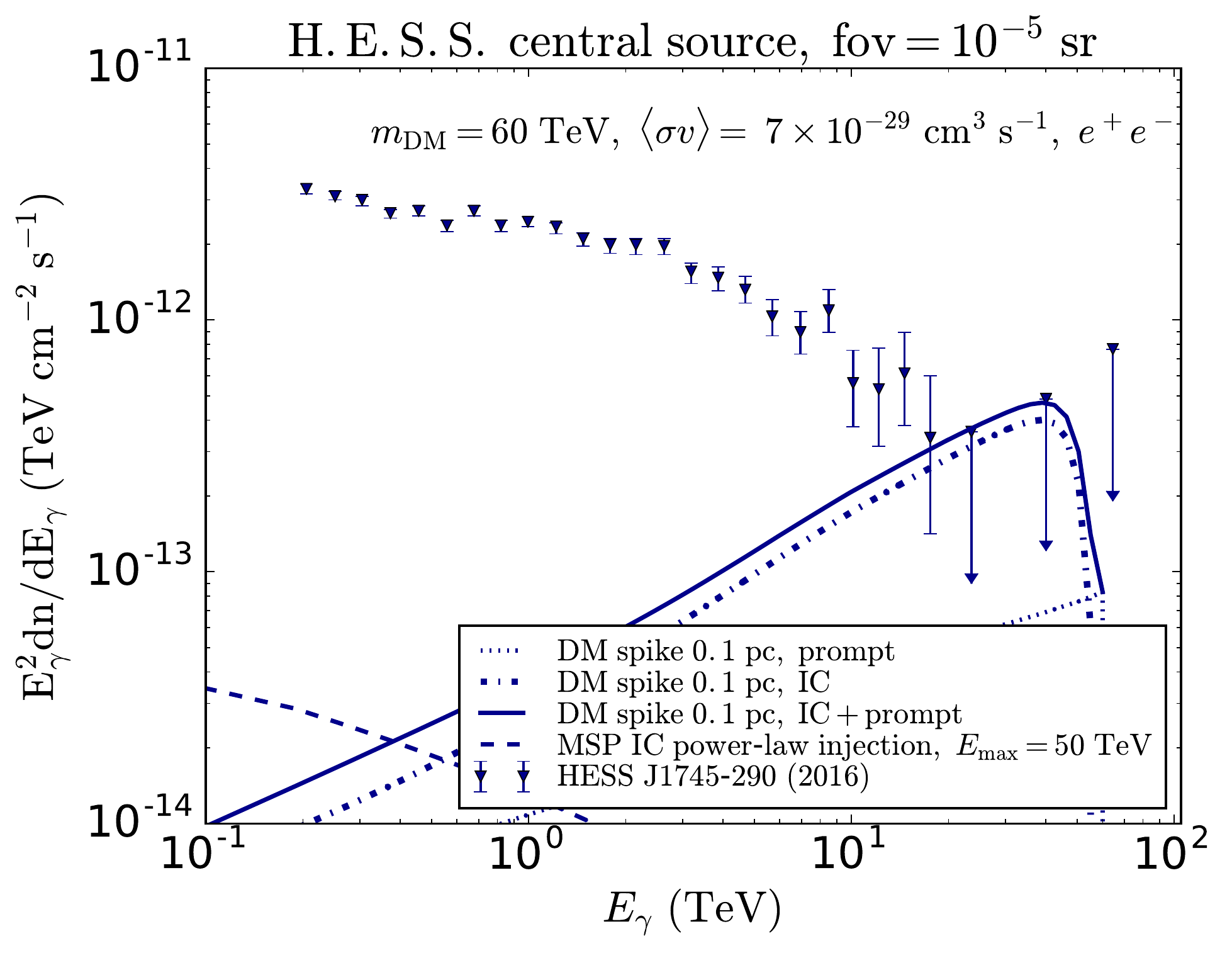}
\caption{\label{extended_e}\textbf{\textsf{Left panel}}: $\gamma$-ray spectra from 100 GeV to 100 TeV for a fov of $1.4 \times 10^{-4}\ \mathrm{sr}$ corresponding to the H.E.S.S. diffuse emission. IC emission from MSPs is depicted as a dashed line. IC emission from a spike of radius 0.1 pc, for a 60 TeV DM candidate annihilating exclusively to $e^+e^-$ with a cross-section of $\left\langle \sigma v \right\rangle = 7 \times 10^{-29}\ \mathrm{cm^{3}\ s^{-1}}$ is shown as a dot-dashed line. The solid line represents the total spectrum. \textbf{\textsf{Right panel}}: $\gamma$-ray spectra from 100 GeV to 100 TeV for a fov of $10^{-5}\ \mathrm{sr}$ corresponding to the central source, HESS J1745-290. In addition to IC emission from MSPs (dashed) and a DM spike of radius 0.1 pc (dot-dashed), the central region features the sharply peaked prompt emission from the spike (dotted). The solid line is the total emission. The MSP and DM parameters are the same as for the left panel. The data points for both panels are taken from Ref.~\cite{HESSextended2016}.}
\end{figure*}

\subsubsection{Inverse Compton emission from DM annihilation}

Prompt emission from a DM spike is the dominant source of DM-induced $\gamma$-rays in the central parsec, but the corresponding spatial extension is too small to account for the H.E.S.S. diffuse emission detected up to 0.45 deg. However, in addition to prompt emission, we expect a significant amount of $\gamma$-rays to arise from IC emission from energetic $e^{\pm}$ produced in DM annihilations. Since $e^{\pm}$ undergo spatial diffusion, the resulting $\gamma$-ray emission can be significantly more spatially extended than the initial DM profile. 

Considering that with our DM spike model, $e^{\pm}$ are produced by DM annihilations below parsec scales, we no longer use the GALPROP ISRF in this case, but the one computed in Ref.~\cite{Kistler:2015yrf} referred hereafter to as the Kistler ISRF, and we assume a smooth reconnection with the larger scale GALPROP ISRF.\footnote{Therefore we consider an effective two-zone model in which the DM component is sensitive to the inner enhanced ISRF, while the shallower MSP distribution is sensitive to the larger scale GALPROP ISRF.} The Kistler enhanced radiation field accounts for the strong photon sources in the central parsec of the Galaxy, and is about three orders of magnitude larger than the GALPROP ISRF. 

We also account for absorption of $\gamma$-rays from $e^+e^-$ pair production on ISRF photons, using the attenuation factor computed in Ref.~\citep{Kistler:2015yrf}. In our model, the attenuation of the $\gamma$-ray emission from pair production on the ambient target photons is increased as shown in Ref.~\citep{Kistler:2015yrf}---given the higher density of the ISRF used here in the central pc---compared to the findings of Ref.~\cite{Moskalenko:2005ng}. However, absorption is still essentially relevant above 10 TeV, and leads to a reduction of the flux of 10\% at $E_{\gamma} \sim 10\ \mathrm{TeV}$, up to 30\% at 100 TeV.

The computation of the IC flux for channel $f$ is similar to the MSP case:
\begin{equation}
\label{IC_flux_DM}
\left. E_{\gamma}^{2}\dfrac{\mathrm{d}n_{f}}{\mathrm{d}E_{\gamma}} \right | _{\mathrm{DM}}^{\mathrm{IC}}= \dfrac{E_{\gamma}}{4\pi} \int_{\Delta\Omega} \! \int_{\mathrm{l.o.s.}} \! j_{\mathrm{DM},f}(E_{\gamma},\vec{x}) \, \mathrm{d}s \, \mathrm{d}\Omega,
\end{equation}
where the IC emissivity reads
\begin{equation}
\label{IC emissivity DM}
j_{\mathrm{DM},f}(E_{\gamma},\vec{x}) = 2 \int_{E_{\gamma}}^{m_{\mathrm{DM}}} \! P_{\mathrm{IC,K}}(E_{\gamma},E_{\mathrm{e}},\vec{x}) \psi_{\mathrm{DM},f}(E_{\mathrm{e}},\vec{x}) \, \mathrm{d}E_{\mathrm{e}},
\end{equation}
and the spectrum accounting for diffusion is given by
\begin{equation}
\label{electron_spectrum_DM}
\psi_{\mathrm{DM},f}(E,\vec{x}) = \dfrac{\kappa_{f}}{b_{\mathrm{K}}(E)} \int_{E}^{E_{\mathrm{max}}} \! \tilde{I}_{\vec{x},\mathrm{DM}}(E,E_{\mathrm{inj}}) \left. \dfrac{\mathrm{d}N_{\mathrm{e},f}}{\mathrm{d}E_{\mathrm{inj}}} \right | _{\mathrm{DM}} \, \mathrm{d}E_{\mathrm{inj}},
\end{equation}
where $ \kappa_{f} = 1/2 \left\langle \sigma v \right\rangle_{f} (\rho_{\odot}/m_{\mathrm{DM}})^{2} $, and $b_{\mathrm{K}}$ is the sum of synchrotron and IC losses in the central pc, where the K subscript stands for the Kistler ISRF. We assume a $B = 10 \ \rm \mu G$ magnetic field in the inner Galactic region. The IC energy loss rate $b_{\mathrm{IC,K}}$ is computed following the procedure of Ref.~\cite{Delahaye2010}, which models the ISRF as a superposition of grey-body spectra. Considering the freedom we have on the poorly constrained diffusion set-up below $\sim 100\ \mathrm{pc}$, to compute the DM spike halo function $\tilde{I}_{\vec{x},\mathrm{DM}}$ we again use the above-mentioned diffusion parameter set-up that avoids superluminic diffusion. The electron injection spectrum $\left. \mathrm{d}N_{\mathrm{e},f}/\mathrm{d}E_{\mathrm{inj}} \right | _{\mathrm{DM}}$ from DM annihilation is taken from Ref.~\citep{Cirelli_cookbook} and includes electroweak corrections, relevant at high energies. $\tilde{I}_{\vec{x},\mathrm{DM}}$ is computed using the method described in Ref.~\citep{Spike_GC_my_paper}, which accounts for the steepness of the source term in the cosmic-ray equation due to the DM spike.

\section{Explaining the H.E.S.S. diffuse emission}
\label{Results}

\subsection{Spectral morphology}

Shown in Fig.~\ref{extended_e} are the $\gamma$-ray spectra for the fov corresponding to region of interest of the H.E.S.S. diffuse emission, $1.4 \times 10^{-4}\ \mathrm{sr}$ (left panel, red) and the central source HESS J1745-290, i.e.~$10^{-5}\ \mathrm{sr}$ (right panel, blue). The H.E.S.S. data points are taken from Ref.~\cite{HESSextended2016}. The dashed and dot-dashed lines represent our predictions of IC emission from MSPs and a DM spike of radius 0.1 pc---corresponding roughly to the size of the gravitational sphere of influence of the central BH---, respectively. Prompt emission from the spike also contributes to the flux in the central source region (right panel, dotted line). We consider a DM candidate of mass $m_{\mathrm{DM}} = 60\ \mathrm{TeV}$, annihilating to $e^+e^-$ with a sub-thermal best-fit cross section of $\left\langle \sigma v \right\rangle = 7 \times 10^{-29}\ \mathrm{cm^{3}\ s^{-1}}$. 

We note that a rather weak annihilation cross-section is required to match the diffuse energy spectrum measured by H.E.S.S.  Such a value can be easily accommodated through thermal p-wave DM annihilations. The rather large DM particle mass is well below the constraint obtained from the unitarity limit \cite{Griest1990} which is well relaxed in case of p-wave annihilations.\footnote{The high mass range has not been extensively explored yet. Models like minimal DM predict DM masses up to a few tens of TeV (see e.g.~Ref.~\cite{DelNobile2015}), but it is possible to go beyond the weak scale, up to very large masses, see e.g.~\cite{Berlin2016,Davoudiasl2016}. However these models correspond so far to soft channels. Still, over the past few years model building has been often motivated by phenomenology, which actually has strong ties with leptonic channels in the context of indirect searches. Our model is therefore phenomenological and its theoretical counterpart is beyond the scope of the paper.}

The left panel of Fig.~\ref{extended_e} shows that the H.E.S.S. diffuse emission can be accounted for by the sum of the IC emission from MSPs and a DM spike, with the lower part of the H.E.S.S. spectrum associated with MSPs, and the high energy part above $\sim 10\ \mathrm{TeV}$ with DM. The reduced chi-squared is $\chi^{2}/\mathrm{d.o.f.} = 23/20 \approx 1.2$, showing the quality of the fit.\footnote{We have 22 data points and 2 free parameters, namely the normalization of the MSP flux and the size of the spike, so 20 degrees of freedom (d.o.f.).} As shown in the right panel of Fig.~\ref{extended_e}, our model is compatible with the observed emission from the central source HESS J1745-290, in particular with the upper limits at the highest energies.

\vspace{-0.5cm}

\subsection{Spatial morphology}

\begin{figure}[th!]
\centering
\includegraphics[width=\linewidth]{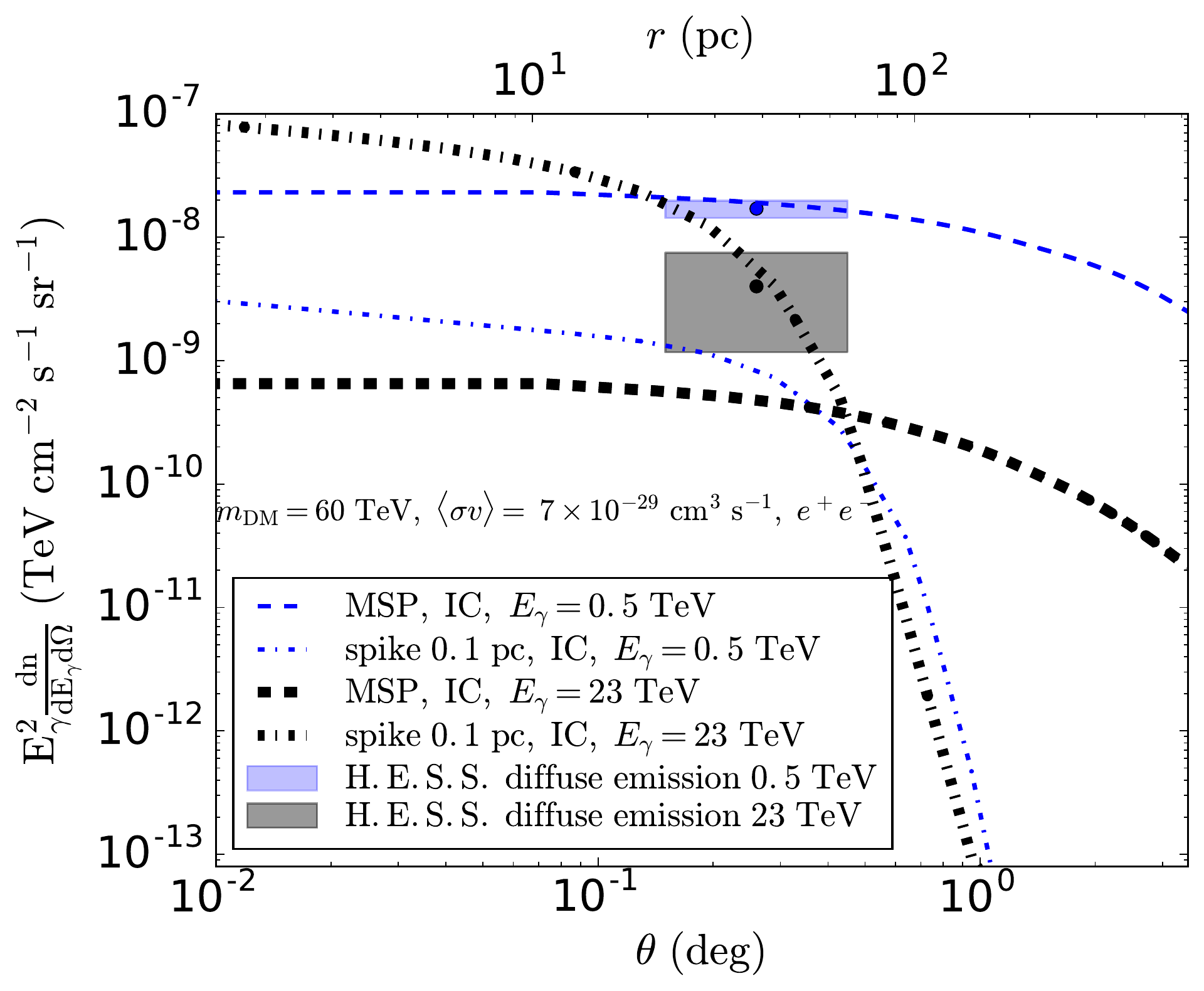}
\caption{\label{spatial}Intensity of IC emission from MSPs (dashed) and a 0.1 pc DM spike (dot-dashed) as a function of angular distance from the GC, at 0.5 TeV (thin blue) and 23 TeV (thick black). The data points at these energies and their statistical error bars are depicted as shaded rectangles.}
\end{figure}

Shown in Fig.~\ref{spatial} are the IC intensities $E_{\gamma}^{2}\mathrm{d}n/(\mathrm{d}E_{\gamma}\mathrm{d}\Omega)$ at 0.5 TeV (thin blue lines) and 23 TeV (thick black lines), as a function of angle $\theta$ (or radius $r$) from the center, for the same components (MSPs, dashed, and DM spike, dot-dashed) as in Fig.~\ref{extended_e}. 

Given the diffusion parameter set-up considered in our model, energetic electrons from DM annihilations can travel out to few 10 pc distance. Fig.~\ref{spatial} shows that for the DM spike, which dominates above $\sim 10\ \mathrm{TeV}$ (see the spectrum in Fig.~\ref{extended_e}), the IC intensity drops steeply around 0.3 deg at 23 TeV and around 1 deg at 0.5 TeV. These specific scales correspond to the diffusion lengths associated with the losses and diffusion coefficient,\footnote{For injection at $\sim 60\ \mathrm{TeV}$ and propagation down to 23 TeV, the diffusion length is $\sim 40\ \mathrm{pc}$ or equivalently $\sim 0.3\ \mathrm{deg}$.} and turn out to be very similar to the characteristic size of the H.E.S.S. diffuse emission. For the MSP component, dominant below $\sim 10\ \mathrm{TeV}$, the spatial extension of the IC emission is of order a few degrees, therefore larger than the H.E.S.S. region. 

The diffuse emission has been detected by H.E.S.S. by accumulating statistics from a significant exposure time in this region. However, the emission might be even more extended, and future H.E.S.S. observations at Galactic latitudes $|b| > 1\ \mathrm{deg} $ would greatly help to discriminate between the proposed scenarios. In particular, according to our predictions, H.E.S.S. should observe an even more extended signal below $\sim 10\ \mathrm{TeV}$, due to the MSP component.

\subsection{Discussion}

First, we point out that we used a smaller magnetic field strength---$10\ \mu\mathrm{G}$ compared to 0.1 mG---and a slightly larger diffusion coefficient at the highest energies than the authors of Ref.~\cite{HESSextended2016}, which accounts for the different conclusions regarding the validity of a leptonic scenario.

The IC flux from the DM spike is sensitive to the losses and diffusion coefficient in the central pc. On the one hand, a magnetic field larger than the $10\ \mu\mathrm{G}$ value we have considered---, for instance a 0.1 mG field \cite{Kistler:2015yrf,Crocker2010,Hinton2007} or a 1 mG field \cite{Nature_mG}, would lead to a significant increase in synchrotron losses, thus significantly reducing the IC flux and spoiling the achievement of explaining the high-energy part of the H.E.S.S. diffuse emission. With a 10 $\mu$G magnetic field, using a milder energy dependence of the diffusion coefficient would only imply a higher normalization of the diffusion coefficient for electrons of a few 10 TeV to diffuse out to the region of interest. However, if the diffusion coefficient was in fact much smaller, typically for Bohm diffusion \cite{Bednarek2013}, the spike-induced IC emission would be confined within the region corresponding to the central source and there would be no leakage into the diffuse emission region. 

Regarding the DM profile, for values of the spike radius $R_{\mathrm{sp}}$ larger than $\sim 0.1\ \mathrm{pc}$, the associated IC flux significantly overshoots both the diffuse and point source data, unless the annihilation cross section is further reduced. Therefore, there is a degeneracy between the cross-section and the spike radius, but this is beyond the scope of this paper. For completeness we computed the IC flux from a heated spike with a 1.5 slope, but the result is roughly two orders of magnitude smaller than the H.E.S.S. flux for the thermal cross section of $3 \times 10^{-26}\ \mathrm{cm^{3}\ s^{-1}}$, so the cross section would have to be increased above observational limits to account for the diffuse emission. Therefore, as mentioned in Sec.~\ref{DM_annihilation}, a SMBH-induced adiabatic spike is required for DM annihilations to account for the high energy part of the H.E.S.S. diffuse emission.

We note that our conclusions depend strongly on the DM annihilation channel, and require dominant annihilation into $e^+e^-$. For softer channels like $\mu^+\mu^-$, $\tau^+\tau^-$ or $b\bar{b}$, the IC flux is too small in the H.E.S.S. extended region of interest while the associated emission in the central 0.1 deg overshoots the flux from the central source HESS J1745-290.

Finally, we also checked that the synchrotron flux from our model does not overshoot the steady diffuse X-ray emission recently detected with the NuSTAR satellite within a few pc of Sgr A*, in the 20--40 keV band \cite{Mori2015,Perez2015}. For a $10\ \mu\mathrm{G}$ magnetic field, the synchrotron flux is actually several orders of magnitude below the measured value.

\section{Conclusion}
\label{Conclusion}

In this study, we have proposed a phenomenological leptonic model of the new diffuse TeV emission observed at the GC with H.E.S.S. that provides a connection between the GeV and TeV scales. More specifically, we have shown that the sum of IC emission from $e^{\pm}$ produced by the same population of MSPs that can explain the \textit{Fermi} GeV excess, and by annihilations of heavy ($\sim 60\ \mathrm{TeV}$) DM particles in a SMBH-induced density spike, can account for the H.E.S.S. diffuse emission. Our model reproduces very well the spectrum of the emission, with MSPs accounting for observations below $\sim 10\ \mathrm{TeV}$ and DM accounting for the higher energy part of the spectrum. We have also discussed the associated spatial morphology. We find that for sensible parameters the DM-induced emission has the same extension as the observed signal, while the size of the MSP component is larger, reaching up to a few degrees. This can be used to test this scenario, depending on whether the current extension is the actual size of the emission region, or if more photon statistics at higher galactic latitudes will uncover a more extended signal.

\acknowledgments{We thank Chris Gordon for fruitful discussions, and the anonymous referee for very useful comments and suggestions. This research has been supported at IAP by the ERC Project No.~267117 (DARK) hosted by UPMC and at JHU by NSF Grant No.~OIA-1124403. This work has also been supported by UPMC and STFC, and has been carried out in the ILP LABEX (ANR-10-LABX-63) and supported by French state funds managed by the ANR, within the Investissements d'Avenir programme (ANR-11-IDEX-0004-02).}

\bibliographystyle{apsrev4-1}  
\bibliography{/Users/thomaslacroix/Documents/thesis/papers/biblio/biblio}

\end{document}